\documentclass[aps,jcp,reprint,twocolumn,citeautoscript,longbibliography,preprintnumbers]{revtex4}
\usepackage{graphicx}
\usepackage[font={small}]{caption}
\usepackage{subcaption}
\usepackage{amsmath}
\usepackage{amssymb}
\usepackage{tabularx}
\usepackage{placeins}
\usepackage{units}
\usepackage{color}
\usepackage{hyperref}
\usepackage[normalem]{ulem}
\usepackage{dcolumn}% Align table columns on decimal point
\usepackage{breakcites}
\usepackage{bm}% bold math
\usepackage{multirow}
\usepackage{adjustbox}
\usepackage{mathtools}
\newcommand{\eref}[1]{Equation~\ref{#1}}
\newcommand{\fref}[1]{Figure~\ref{#1}}
\newcommand{\tref}[1]{Table~\ref{#1}}

\begin{document}

\renewcommand{\arraystretch}{1.2}

\title{ Correlation consistent effective core potentials for late 3d transition metals 
adapted for plane wave calculations}

%\thanks{{\bf Actually we do not need this since there is no coauthor from the lab.} Notice: This manuscript has been authored by UT-Battelle, LLC, under
%contract DE-AC05-00OR22725 with the US Department of Energy (DOE).
%The US government retains and the publisher, by accepting
%the article for publication, acknowledges that the US
%government retains a nonexclusive, paid-up, irrevocable, worldwide
%license to publish or reproduce the published form of this manuscript,
%or allow others to do so, for US government purposes.
%DOE will provide public access to these results
%of federally sponsored research in accordance with the DOE Public
%Access Plan (\url{http://energy.gov/downloads/doe-public-access-plan}).
%}

\author{Benjamin Kincaid$^{1, \dagger}$}
\email{bekincai@ncsu.edu }
\author{Guangming Wang$^{1, \dagger}$}
\author{Haihan Zhou$^1$, and Lubos Mitas$^1$}

\affiliation{
1) Department of Physics, North Carolina State University, Raleigh, North Carolina 27695-8202, USA 
}

\thanks{These authors contributed equally to this work}

\begin{abstract}
We construct a new modification of correlation consistent 
effective potentials (ccECPs) for late $3d$ elements  Cr-Zn with Ne-core
that are adapted for efficiency and low energy cut-offs in plane wave calculations.
The decrease in accuracy is rather minor so that the constructions are in the same overall accuracy class as the original ccECPs.
The resulting new constructions 
work with energy
cut-offs at or below
$\approx$ 400 Ry and thus make calculations of large systems with transition metals feasible for
plane wave codes.
We provide also the basic benchmarks for atomic spectra and molecular tests of this modified option that we denote as ccECP-soft.
\end{abstract}

\maketitle

\section{Introduction}

Key properties of matter such as cohesion, magnetic or  optical responses can be derived from valence electronic structure calculations. Fortunately, electronic levels in atoms show a significant distinction between core and valence states so that it is possible to introduce valence-only effective Hamiltonians.
The fact that core and valence states occupy different ranges both in spatial and in energy domains enables us to partition the atomic states into core and valence subspaces. 
The theory of these well-known pseudopotentials or effective core potentials (ECPs) has been perfected over the decades and it includes a number of criteria that model the
influence of the core on valence electrons as closely as possible to the original, all-electron atom. This includes concepts 
such as norm-conservation of one-particle states,  consistency of energy differences for atomic excitations, different forms and beyond \cite{fernandez_pacios_ab_1985, hurley_1986, lajohn_1987, ross_1990, stevens_relativistic_1992, troullier_efficient_1991, burkatzki_energy-consistent_2007-2, burkatzki_energy-consistent_2008, bergner_ab_1993,dolg_energyadjusted_1987, bachelet_pseudopotentials_1982, trail_shape_2017, vanderbilt_soft_1990, dolg_relativistic_2012, goedecker1998, goedecker2013}.
%\textcolor{red}{GM: cite all these papers here?} {\color{blue} LM: Yes, I actually want to add a few more.  Check the names in Ref.16 }  
The advantages of using ECPs are both qualitative and quantitative.
On the quantitative side, the valence energy scale and number of degrees of freedom are essentially unchanged   across the periodic table with resulting orders of magnitude speed-ups in calculations. On the qualitative front, ECPs can be constructed to mimic true atoms by effectively taking into account not only single-particle picture but also core-core correlations, impact of core-valence correlations, spin-orbit and other relativistic effects
by employing surprisingly simple, nonlocal, 
one-particle operators
with $lm-$ or $jlm$-projectors. Of course, there is a price for these advantages since there is always some ECP-related bias present. However, over the years, the accuracy of high quality ECPs has been steadily improving and, at present, the corresponding errors typically do not dominate electronic structure calculations.
%by being typically below chemical or even sub-chemical accuracy thresholds.  

Recently, we have introduced a new generation of such effective Hamiltonians which we call correlation consistent ECPs (ccECPs)
\cite{bennett_new_2017}.
We have emphasized several key principles in order to significantly improve the fidelity with respect to the original,
all-electron Hamiltonian
in broad classes of calculations such as molecular systems, condensed matter materials with periodic or mixed boundary conditions.
These principles involve:
i) constructions that minimize discrepancies
between all-electron and ccECP many-body atomic spectra as well as one-particle properties such as charge norm conservations; ii) use many-body
methods that include 
Coupled Cluster in constructions as well
as quantum Monte Carlo (QMC) in testing and benchmarking;
iii) for certain elements we introduced several sizes of valence and core subspaces including all-electron regularized Coulomb  potentials for very light elements (for example, reg-ccECPs for H-Be elements);
iv) easy use with simple parametrizations in gaussian expansions; v) 
open data website with full access and further adapting to particular types of calculations.

The derived ccECPs indeed turned out to be, in general, more consistently accurate than previous constructions and they also provide better balance of accuracy in various settings. In particular, the tests on molecules in non-equilibrium geometries have demonstrated significantly improved transferability apart from equilibrium atomic conformations. Some deviations from chemical accuracy have occurred for early main group elements in $3s3p$ and $4s4p$ columns at very short bond lengths of oxide molecules. This is a 
well-known limitation due to the small number of valence electrons and polarizability of the most shallow core states. For $3s3p$ elements this issue has been addressed by 
providing a [He]-core option that makes the calculations essentially equivalent to the all-electron setting. Being well-defined and tested,
ccECPs also enabled us to put a bound on systematic biases in quantum Monte Carlo calculations,
 for example, in the study of molecular and solid state systems
 including very large supercell sizes with hundreds of valence electrons
\cite{wang_new_2019,annaberdiyev_cohesion_2021}.
 Therefore, ccECPs provide both practical tool as well as a valuable accuracy standard for the benchmarking of other constructions and as such they are
being independently probed by the community at large \cite{zhouDiffusionQuantumMonte2019}.
%It is interesting that for several elements it is possible to find almost equally accurate previously constructed ECPs for some quantities, however, overall accuracy and balance appears essentially always better or, at worst, on par with ccECPs.

At the same time, the higher accuracy implies deeper potential functions that make the valence-only electronic states less smooth and more curved in the core region. 
This is fundamentally correct since it mimics more accurately the
distribution of effective valence charges inside the core. This is corroborated by capturing
the correct shape of molecular binding curves for hydride and oxide dimers that probe both covalent and polarized bonds. For most elements the conventional core ccECPs can be 
used both in plane wave calculations with energy cut-offs below roughly 400 Ryd
with converged energy to
 1 meV/electron or so. Unfortunately,
for late $3d$ transition metals with 
a very deep $3s$, $3p$ semicore and localized $3d$ states, the cost of plane wave calculations goes up very significantly, needing a cut-off of around 1400 Rydberg or more.%due to needed cut-off of, say, 1600 Ryd or more. 
%\textcolor{red}{BK: I think the end of the previous sentence reads a little rough. I'll revise it, but keep what was written as a comment in case we want to keep it as is.}%\cite{XXXX}.
%\textcolor{red}{which paper to cite} {\color{blue} LM: Good question, I actually do not remember any. Maybe we will just write it qualitatively, without reference.}

In order to overcome this limitation we have identified possible ways to decrease the energy cut-offs while keeping the accuracy at the level comparable to the original ccECPs. 
%The resulting significant gains 
%in speed-ups due to much lower plane wave cutoffs make them comparable to other ccECPs such as those in the $2s2p$ row. 
Note that there is often a very subtle 
balance between the parameterization details on one hand and their impact 
on ECP properties on the other.
%that has to be balanced with acceptable resulting properties.
As we commented upon before
\cite{bennett_new_2017}, even rather minor changes in ECP parameters could push either the accuracy, charge smoothness
or curvature 
%close to the ion 
in directions that are counterproductive. Similarly, over constraining of cost (objective) functions
and/or over parameterizations could be counterproductive as well since that could
lead to linear dependencies, undue costly optimizations and overall inefficiency. 
In what follows we 
briefly describe the results of these adapted constructions, the corresponding 
forms, relevant updates of previously introduced methods, results and testing.
The resulting ccECP-soft constructions exhibit much shallower potential functions and offer major efficiency gains in plane wave calculations.
%{\bf expand this here with longer description}. 

\section{ECP Form}
The parameterization of the ECPs are unchanged from \cite{annaberdiyev_new_2018}, following a semi-local format. 
\begin{equation}
    V_{i} ^{pp} = V_{loc}(r_i) + \sum _{\ell = 0} ^{\ell_{max}} V_{\ell}(r_i) \sum_m |\ell m\rangle \langle \ell m|
    \label{eq:ecp_parameterization}
\end{equation}
Where $r_i$ indicates the radial contribution of the $i^{th}$ electron, and $\ell_{max}$ is 1 for the elements investigated in this work. $V_{loc}$ is chosen to cancel out the coulomb singularity. $V_{loc}(r)$ has the form:
\begin{equation}
    V_{loc}(r) = - \frac{Z_{\rm eff}}{r}(1-e^{-\alpha r^2}) + \alpha Z_{\rm eff} r e^{-\beta r ^2} + \sum_{i=1} ^{2}\gamma_i e^{-\delta_i r^2} 
    \label{eq:local_channel}
\end{equation}
where $Z_{\rm eff}$ is the effective charge of the valence space
%, e.g. the charge, $Z$, of the element subtracting the charge of the fixed core, 
while $\alpha$, $\beta$, $\delta_i$, and $\gamma_i$ are parameters determined by the optimization. This form explicitly cancels the Coulomb singularity and insures smooth behavior at very small radii \cite{burkatzki_energy-consistent_2008}.   

The non-local potentials are parameterized as
\begin{equation}
    V_{l}(r) = \sum_{j=1}^{2}\beta_{\ell j} r^{n^{\ell j} - 2} e^{-\alpha_{\ell j} r^2}
\end{equation}
where $\beta_{\ell j}$ and $\alpha_{\ell j}$ are optimized for each non-local channel.

\section{Method Updates}
The methods employed in this work are largely adapted and updated from our previous works\cite{annaberdiyev_new_2018}; therefore, here we recount only the basic points and current modifications. 
Most of the differences from the previous publication come in the form of more stringent constraints during the optimization. 
In the cited work, the exponents of the gaussian expansions that form the pseudopotential were allowed to freely move during optimization, thus they could increase or decrease as needed within a very wide range of permitted values. 
Since the original objective was to maximize accuracy and fidelity when compared to the all-electron setting, the resulting pseudopotential functions were in general varying so as to reflect the true all-electron and bare ion interactions experienced by a valence electron.

In order to construct softer ECPs, constraints were placed on the maximum allowed values for the exponents. 
This has two effects. 
First, it  smooths out the curvature of the potentials and, second, it indirectly restricts the potentials' amplitudes. The resulting potentials are therefore more shallow and exhibit lower curvatures. This in turn leads to lower cutoffs in plane wave expansions. 
Lastly, in our previous research we included highly ionized states in the objective function to construct each ccECP, while in this work we only ionize to around $+4$ or $+5$. This focuses the objective function to only account for likely configurations that could feasibly occur in molecules and solids and makes the optimization easier to perform reliably. 
%\textcolor{red}{BK: I added the last couple of lines to explain another difference with this round of ECPs. I can refine the wording if we think it's necessary.}
Thus the optimization protocol is as follows:
\begin{enumerate}
    \item Calculate high accuracy CCSD(T) all-electron (AE) data for each element that involves a set of excited states within the desired energy window, and generate initial ECP candidates with confirmed cutoffs below the desired threshold.
    \item Applying the same techniques we previously employed\cite{annaberdiyev_new_2018}, incorporate correlation energy into the optimization implicitly by finding the contributions to the energy from the scalar relativistic Dirac-Fock calculation and then correlation contributions from CCSD(T) calculations. 
    Using the AE data the gaps are shifted as: $\Delta E_{shifted} = \Delta E_{AE} - 
    \Delta E_{ECP}^{corr}$ where $\Delta E_{AE}$ is the AE CC gap, and $\Delta E_{ECP}^{corr}$ is the ECP's correlation energy contribution for the same gap. 
    This is viable because the correlation energy between ECPs that share the same valence space tends to remain largely constant as shown in our previous paper\cite{annaberdiyev_new_2018}. 
    These new gaps will serve as the major component of the objective function. 
    \item Considering a many-body spectrum $S$, the objective function $\Gamma$ is given by 
    \begin{align}
        \Sigma(S) &= \sum _{s \in S} w_s (\Delta E_{ECP}^{(s)} - \Delta E_{shifted}^{(s)})\\
        \Gamma &= \Sigma(S) + \gamma\sum _{\ell} ( \epsilon_{\ell} ^{ECP} - \epsilon_{\ell} ^{AE} )^2
        \label{eq:obj_func}
    \end{align}
    and it is 
    minimized from an initial guess using the DONLP2 routine \cite{donlp2}.
    $\epsilon_{\ell}^{AE}$ denote the AE one-particle scalar relativistic Hartree-Fock eigenvalues for the semi-core orbitals ($3s$ and $3p$ in this case) and similarly $\epsilon_{\ell}^{ECP}$ denote the corresponding Hartree-Fock eigenvalues. 
    \item Step 3 will keep iterating to minimize the objective function, resulting in the the final ECP once a sufficient minimum is reached. 
    
\end{enumerate}
\eref{eq:obj_func} is the same as seen in previous work by the group on the $3d$ transition metals\cite{annaberdiyev_new_2018}. 
As these optimizations follow a multi-variate minimization scheme and the objective function landscape exhibits multiple valleys,  the quality of the final output depends on the initial guess and it also optimization procedure dependent. In order to overcome this, we have used a number of different initial starting points as well  
as different sets of weights that enable the routine to explore a much larger part of the parameter space. This has provided
well-optimized solutions that fulfill the imposed accuracy requirements. 

The overall target for the plane wave energy cut-off was
$\approx 400$ Ry. We found this value to be close to the "sweet spot" with regard to the balance between accuracy vs gains in efficiency. Of course, this value serves only as a guiding parameter since actual cut-offs in various codes will depend on calculated systems, accuracy criteria, etc. %makes the optimization more robust and allows the optimizer to avoid sub-optimal local minimas in the parameter space.

\section{Results}

All the ECP parameters are given in \tref{tab:ecp_params}.
For all updated ECPs in this work, the core removed is the innermost 10 electrons, referred to as a [Ne]-core for simplicity. 

The errors of the atomic spectrum for ccECP 
and ccECP-soft are evaluated by the mean 
absolute deviation (MAD) of considered atomic excitations
that include bounded anions, $s\to d$ transfers and cations up to 3rd/4th ionizations
\begin{equation}
\label{eqn:mad}
    \mathrm{MAD} = \frac{1}{N} \sum_{i}^{N} \left| \Delta E_{i}^{\textrm{ECP}} - \Delta E_{i}^{\textrm{AE}} \right|.
\end{equation}
For all the elements, the errors are provided at complete basis limit (CBS). The CBS atomic state energies are estimated from the extrapolation method we used in our previous papers. \cite{bennett_new_2018, annaberdiyev_new_2018, wang_new_2019} The detailed data of the AE spectrum and ECP discrepancies for each atom
can be found in the supplementary material.

In \tref{fig:MAD_in_elements}, we show the summary of spectral errors of ccECP and ccECP-soft compared to AE calculations.
ccECP-soft constructions show mildly larger discrepancies compared to ccECPs, but the  MAD errors are essentially within chemical accuracy for all the elements. 
All energies involved are in eV unless specified otherwise.

\begin{table*}%[htbp!]
\small
\centering
\caption{Parameters for the ccECP-soft. For all ECPs, the highest $\ell$ value corresponds to the local channel $L$.
Note that the highest non-local angular momentum channel $\ell_{max}$ is related to it as $\ell_{max}=L-1$.
}

\label{tab:ecp_params}
\begin{adjustbox}{width=0.64\textwidth, center}
\begin{tabular}{ccrrrrcccrrrrr}
\hline\hline
\multicolumn{1}{c}{Atom} & \multicolumn{1}{c}{$Z_{\rm eff}$} &  \multicolumn{1}{c}{$\ell$} & \multicolumn{1}{c}{$n_{\ell k}$} & \multicolumn{1}{c}{$\alpha_{\ell k}$} & \multicolumn{1}{c}{$\beta_{\ell k}$} & & \multicolumn{1}{c}{Atom} & \multicolumn{1}{c}{$Z_{\rm eff}$}  & \multicolumn{1}{c}{$\ell$} & \multicolumn{1}{c}{$n_{\ell k}$} & \multicolumn{1}{c}{$\alpha_{\ell k}$} & \multicolumn{1}{c}{$\beta_{\ell k}$} \\
\hline
Cr & 14 &  0 & 2 &    9.800322 &   89.846846   && Mn & 15 &  0 & 2 &   11.244397 &   57.880958  \\
   &    &  0 & 2 &    8.010010 &   18.997257   &&    &    &  0 & 2 &   11.614251 &   92.965750  \\
   &    &  1 & 2 &    8.785958 &   44.926062   &&    &    &  1 & 2 &    8.702628 &   44.447892  \\
   &    &  1 & 2 &    7.014726 &   14.003861   &&    &    &  1 & 2 &   14.217018 &   41.889380  \\
   &    &  2 & 1 &    3.497383 &   14.000000   &&    &    &  2 & 1 &    4.039945 &   15.000000  \\
   &    &  2 & 3 &    3.611831 &   48.963362   &&    &    &  2 & 3 &    4.200000 &   60.599175  \\
   &    &  2 & 2 &    3.449201 &  -56.466431   &&    &    &  2 & 2 &    4.139297 &  -65.806234  \\
   &    &  2 & 2 &    2.009794 &    0.968440   &&    &    &    &   &             &              \\
   &    &    &   &             &               &&    &    &    &   &             &              \\
Fe & 16 &  0 & 2 &   13.221833 &  153.088061   && Co & 17 &  0 & 2 &   11.423427 &   90.855286  \\  
   &    &  0 & 2 &    7.769539 &   11.680385   &&    &    &  0 & 2 &    9.920127 &   25.185194  \\  
   &    &  1 & 2 &    9.100629 &   40.685923   &&    &    &  1 & 2 &    9.811352 &   48.270556  \\  
   &    &  1 & 2 &    7.483933 &   14.200485   &&    &    &  1 & 2 &    9.340854 &   14.620602  \\  
   &    &  2 & 1 &    3.798917 &   16.000000   &&    &    &  2 & 1 &    3.932921 &   17.000000  \\  
   &    &  2 & 3 &    3.576729 &   60.782672   &&    &    &  2 & 3 &    4.547187 &   66.859664  \\  
   &    &  2 & 2 &    3.514698 &  -66.518840   &&    &    &  2 & 2 &    4.242934 &  -76.154505  \\  
   &    &  2 & 2 &    3.058692 &    1.621670   &&    &    &  2 & 2 &    2.106360 &    1.483219  \\  
   &    &    &   &             &               &&    &    &    &   &             &              \\
Ni & 18 &  0 & 2 &   10.199961 &   41.053383   && Cu & 19 &  0 & 2 &   12.068348 &   78.019159  \\
   &    &  0 & 2 &   11.552726 &   66.727192   &&    &    &  0 & 2 &    9.360313 &   27.107011  \\
   &    &  1 & 2 &    8.131870 &   24.281961   &&    &    &  1 & 2 &   13.173488 &   54.905280  \\
   &    &  1 & 2 &   11.380045 &   36.306696   &&    &    &  1 & 2 &    6.969207 &   14.661758  \\
   &    &  2 & 1 &    3.641646 &   18.000000   &&    &    &  2 & 1 &    3.806452 &   19.000000  \\
   &    &  2 & 3 &    3.641643 &   65.549624   &&    &    &  2 & 3 &    4.021416 &   72.322595  \\
   &    &  2 & 2 &    3.637271 &  -73.527489   &&    &    &  2 & 2 &    3.885376 &  -84.688200  \\
   &    &  2 & 2 &    3.327582 &  -0.856416    &&    &    &  2 & 2 &    2.626437 &    3.393685  \\
   &    &    &   &             &               &&    &    &    &   &             &              \\
Zn & 20 &  0 & 2 &   12.006960 &   56.869394   &&    &    &    &   &             &              \\
   &    &  0 & 2 &    9.103589 &   34.859484   &&    &    &    &   &             &              \\
   &    &  1 & 2 &   10.245529 &   32.153902   &&    &    &    &   &             &              \\
   &    &  1 & 2 &    7.286335 &   15.898530   &&    &    &    &   &             &              \\
   &    &  2 & 1 &    3.465445 &   20.000000   &&    &    &    &   &             &              \\
   &    &  2 & 3 &    3.528420 &   69.308902   &&    &    &    &   &             &              \\
   &    &  2 & 2 &    3.545575 &  -83.673652   &&    &    &    &   &             &              \\
   &    &  2 & 2 &    2.234272 &    0.840046   &&    &    &    &   &             &              \\
%   &    &    &   &             &               &&    &    &    &   &             &               \\
\hline\hline
\end{tabular}
\end{adjustbox}
\end{table*}

\begin{figure}[!htbp]
\centering
\includegraphics[width=0.90\columnwidth]{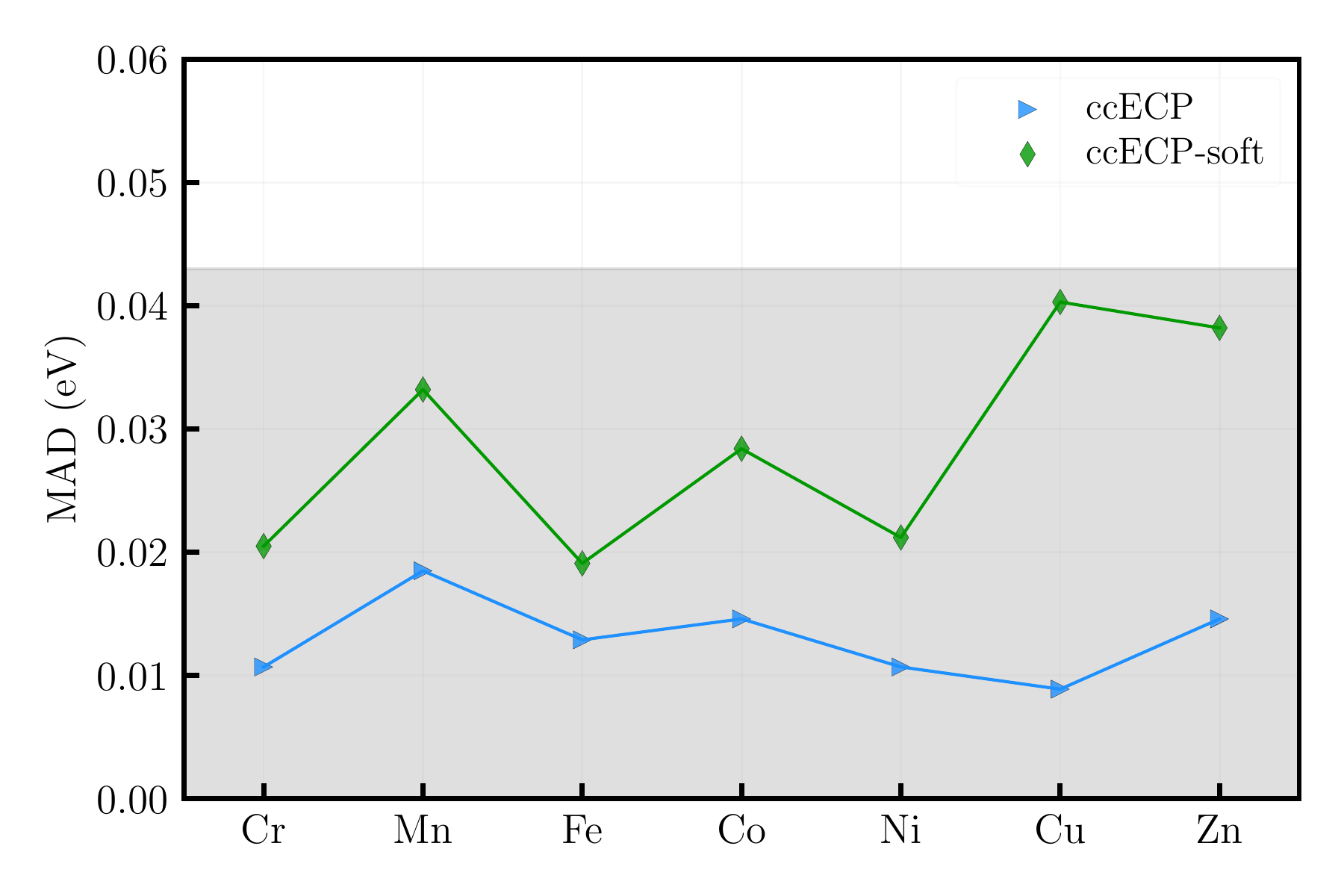}
\caption{ Mean absolute deviations (MADs) of atomic excitations for ccECP and ccECP-soft with the reference represented by the scalar relativistic RCCSD(T) method.
}
\label{fig:MAD_in_elements}
\end{figure}

\subsection{Cr}
\begin{figure*}[!htbp]
\centering
\begin{subfigure}{0.5\textwidth}
\includegraphics[width=\textwidth]{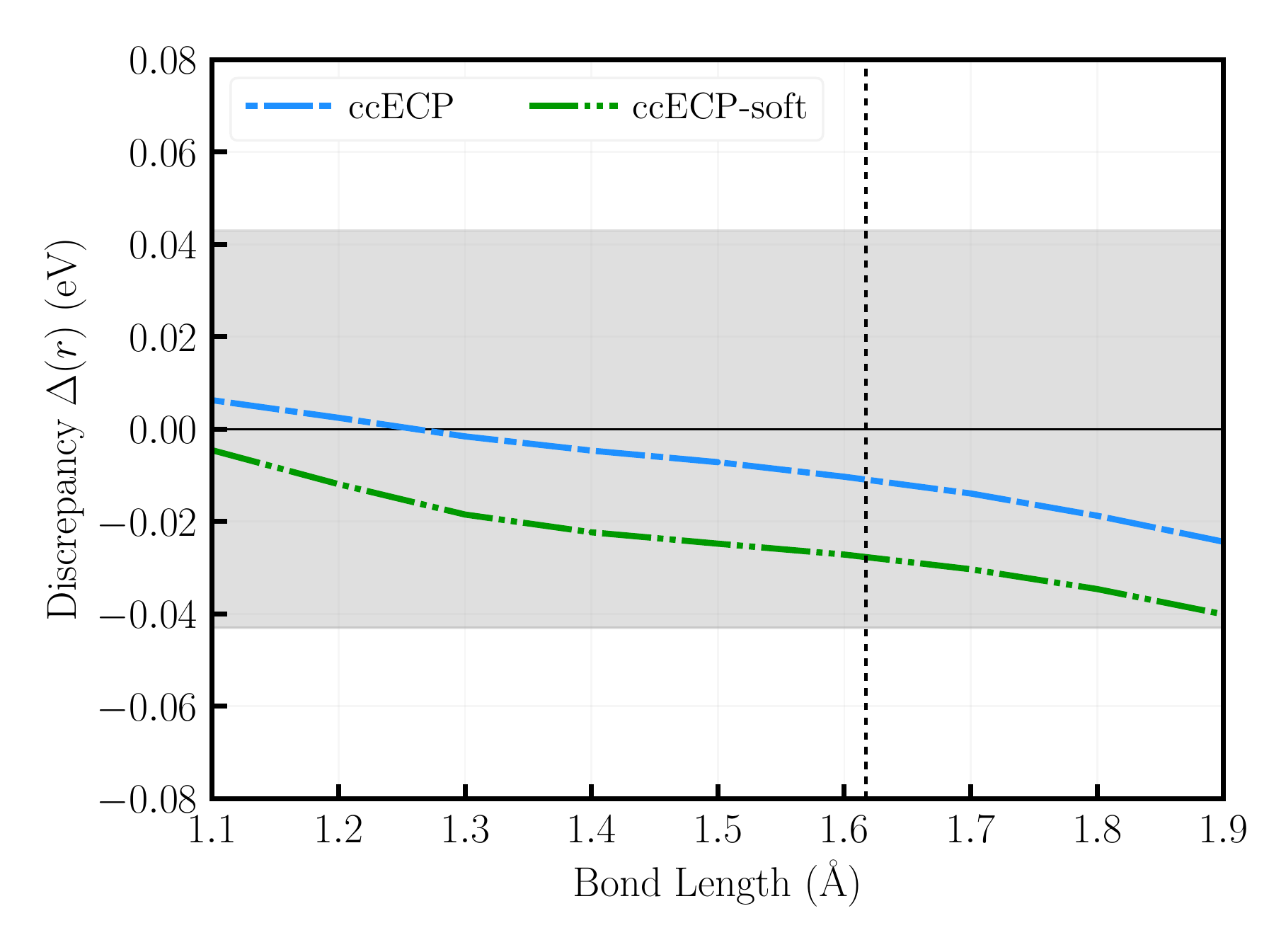}
\caption{CrH 5Z binding curve discrepancies}
\label{fig:CrH}
\end{subfigure}%
\begin{subfigure}{0.5\textwidth}
\includegraphics[width=\textwidth]{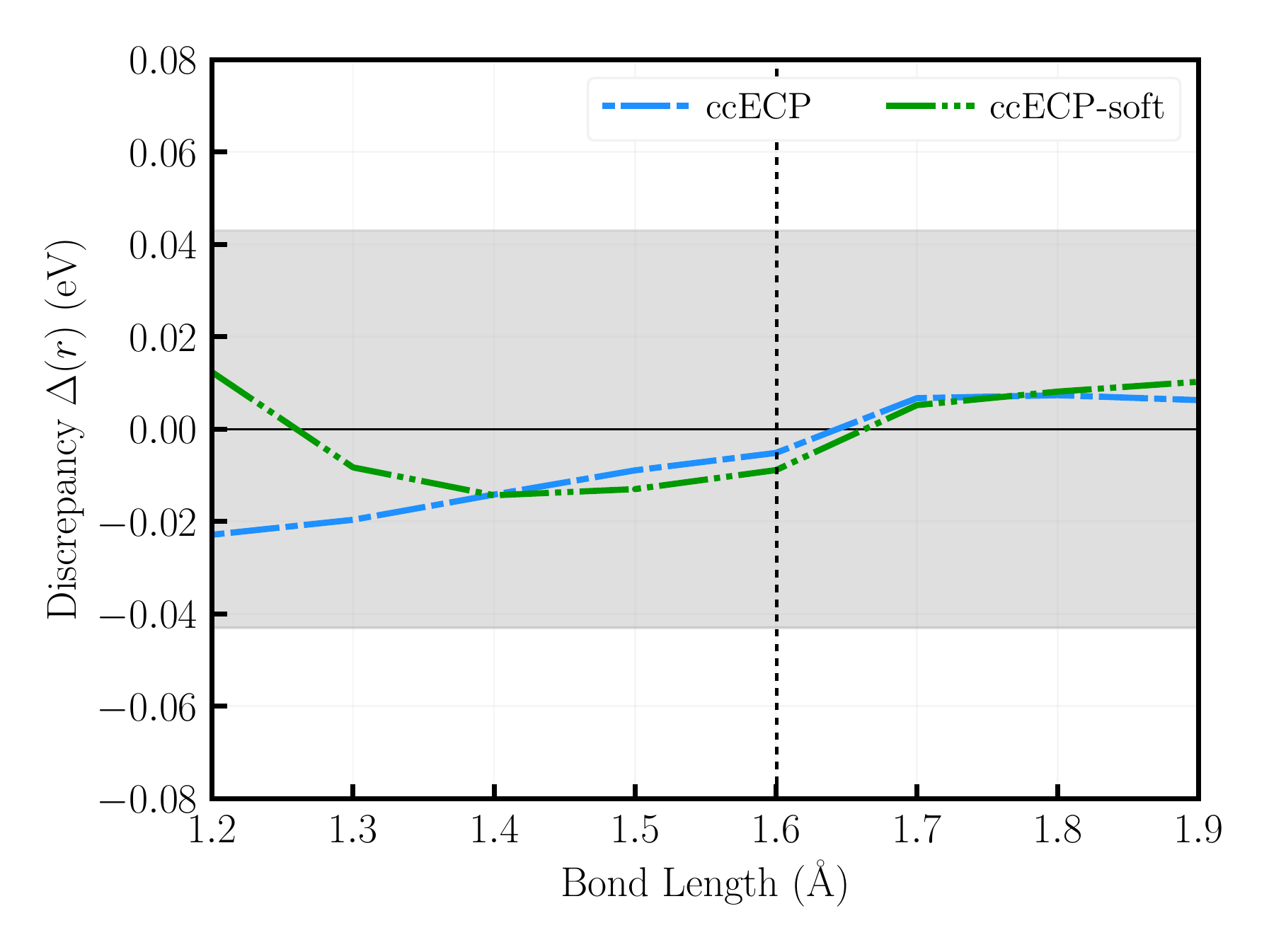}
\caption{CrO QZ binding curve discrepancies}
\label{fig:CrO}
\end{subfigure}
\caption{
Binding energy discrepancies for (a) CrH and (b) CrO molecules with the reference being the scalar relativistic all-electron CCSD(T) result.
The shaded region indicates the band of chemical accuracy. The dashed vertical line represents the equilibrium geometry.
}
\label{fig:Cr_mols}
\end{figure*}

Cr starts off the set and showcases some of the compromises made to soften the previously made ccECPs. The spectrum performance is of a similar quality for the chosen spectra sets as the standard ccECP, but is decidedly worse at higher ionizations. 
In general all of the ccECP-soft pseudopotentials struggle to capture the nature of highly ionized states and as such required truncated training sets compared to the standard ccECPs. 
Despite this, Cr does well to demonstrate the benefits that such compromises can lead to. The highest discrepancy observed in the spectral states was $0.042$eV for the $[Ar]d^6$ state, confirming that all states tested lie within chemical accuracy for the spectrum. 
Similarly, the binding energy curves for CrH and CrO in \fref{fig:Cr_mols} shows that the discrepancies lie within chemical accuracy across all of the geometries investigated. 
Thus, with the minor compromise on the overall quality of the ccECP, we were able to specialize the ccECP-soft version with significant decrease of plane wave cutoff to about 300 Ry.
%Thus, without compromising much on the overall quality of the ccECP, we were able to specialize this version for use in planewave codes.  

\subsection{Mn}
%The valence space for the Mn ECP is $3s^2 3p^6 4s^2 3d^5$. 
The new ccECP-soft construction for Mn has a slightly different from from the rest, possessing only 3 gaussians in the local channel as opposed to 4 used for the rest of the ECPs in the series. However, the spectrum shows a very good agreement with the all electron values and the largest error of $\approx$ 0.06 eV for $^4F(3d^7)$ state and very good balance for the rest of states.  Similarly,
the molecular curves are within the chemical accuracy.
%\textcolor{blue}{Should I begin preparing a new version that falls in line with the rest after this first draft is ready?}

\begin{figure*}[!htbp]
\centering
\begin{subfigure}{0.5\textwidth}
\includegraphics[width=\textwidth]{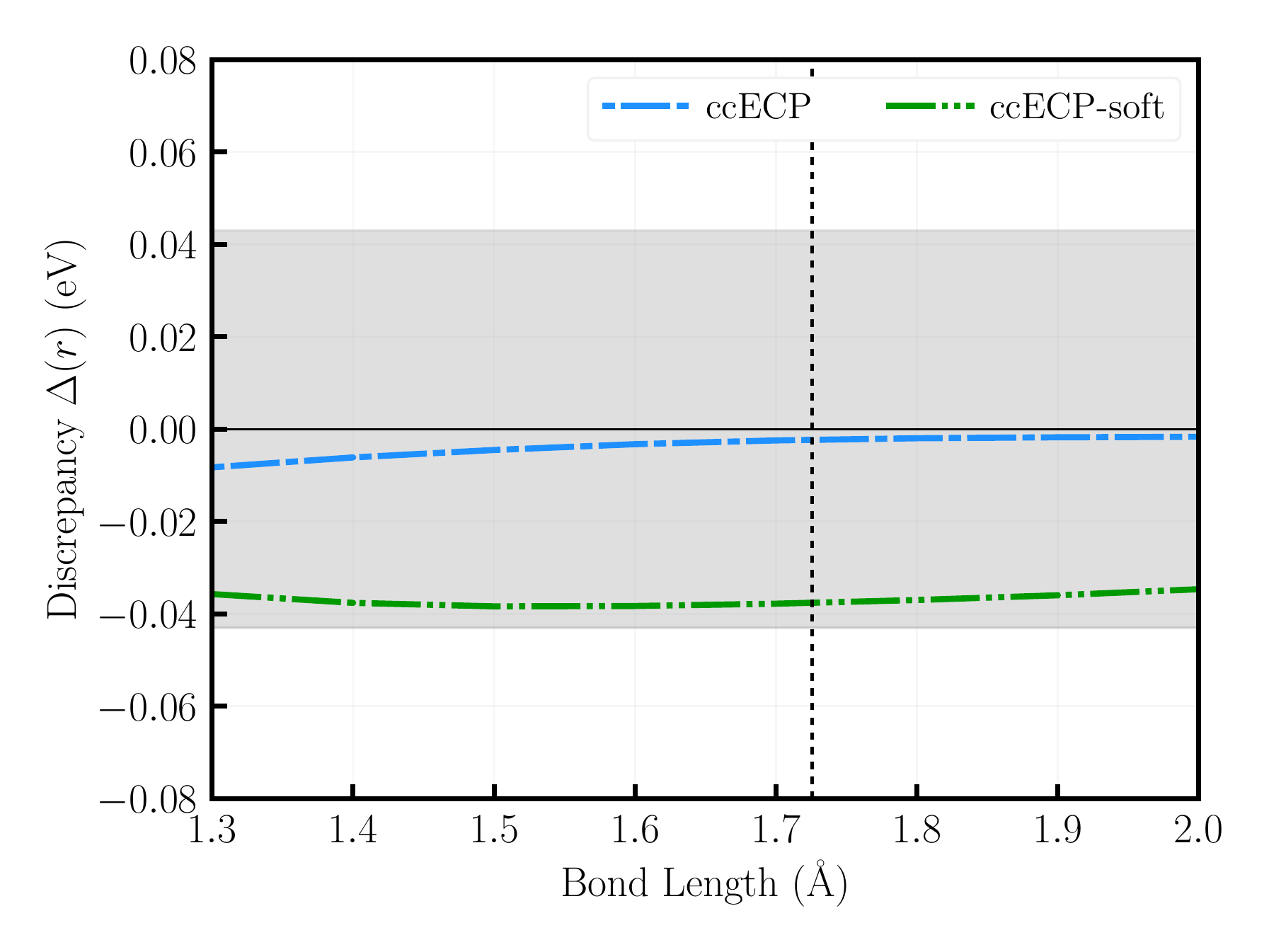}
\caption{MnH 5Z binding curve discrepancies}
\label{fig:MnH}
\end{subfigure}%
\begin{subfigure}{0.5\textwidth}
\includegraphics[width=\textwidth]{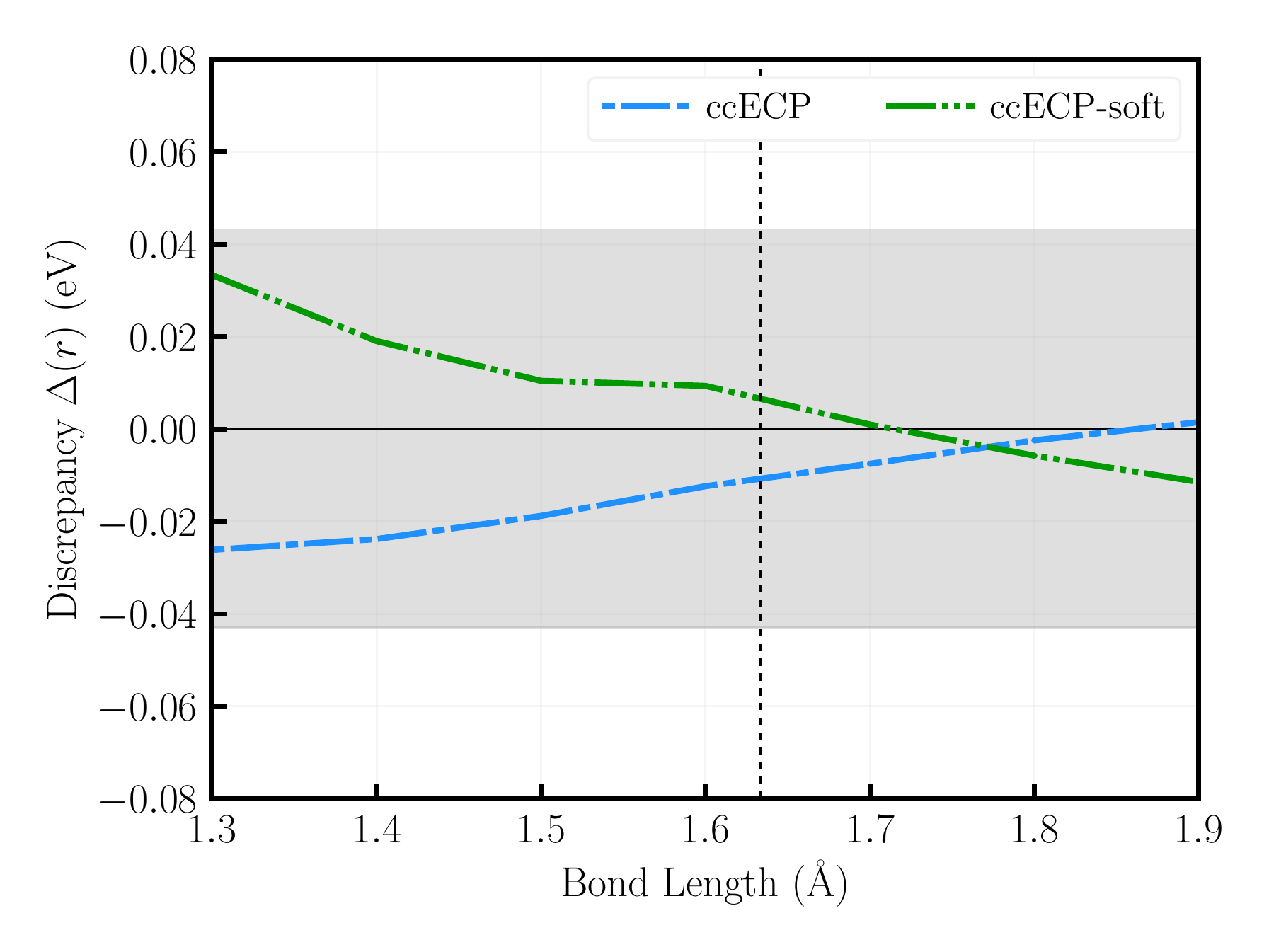}
\caption{MnO QZ binding curve discrepancies}
\label{fig:MnO}
\end{subfigure}
\caption{
Binding energy discrepancies for (a) MnH and (b) MnO molecules with the reference being the scalar relativistic all-electron CCSD(T) result.
The shaded region indicates the band of chemical accuracy. The dashed vertical line represents the equilibrium geometry.
%\textbf{GM: Hi Ben, could you please replot these graphs and label the lines correctly here?}
}
\label{fig:Mn_mols}
\end{figure*}

\subsection{Fe}

For the Fe atom we achieved 
very convincing results, our ccECP-soft version is almost fully comparable to original  ccECP in overall accuracy. For comparison purposes, we also show molecular discrepancies from the DFT-derived set of
alternative ECPs by Krogel-Santana-Reboredo which are constructed to have
very small cut-offs in general \cite{krogel_pseudopotentials_2016}. 
In order to do so we refitted the 
original form with gaussians so that
we could analyze its behavior and quantify the energy discrepancies 
using  gaussian-based codes. 
For the oxide dimer 
we see noticeable overbinding
bias which increases 
with shortening of the bond length.

\begin{figure*}[!htbp]
\centering
\begin{subfigure}{0.5\textwidth}
\includegraphics[width=\textwidth]{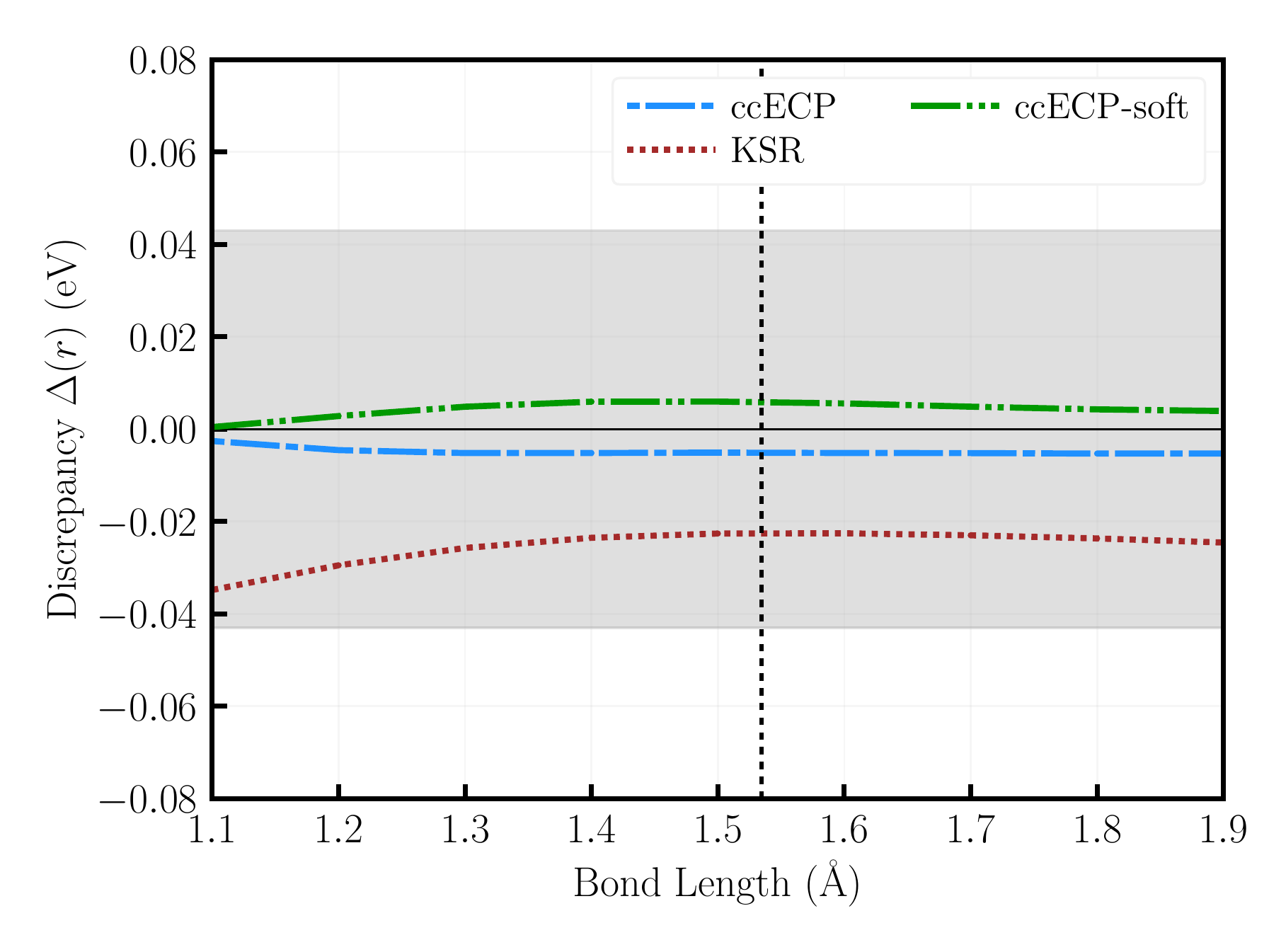}
\caption{FeH 5Z binding curve discrepancies}
\label{fig:FeH}
\end{subfigure}%
\begin{subfigure}{0.5\textwidth}
\includegraphics[width=\textwidth]{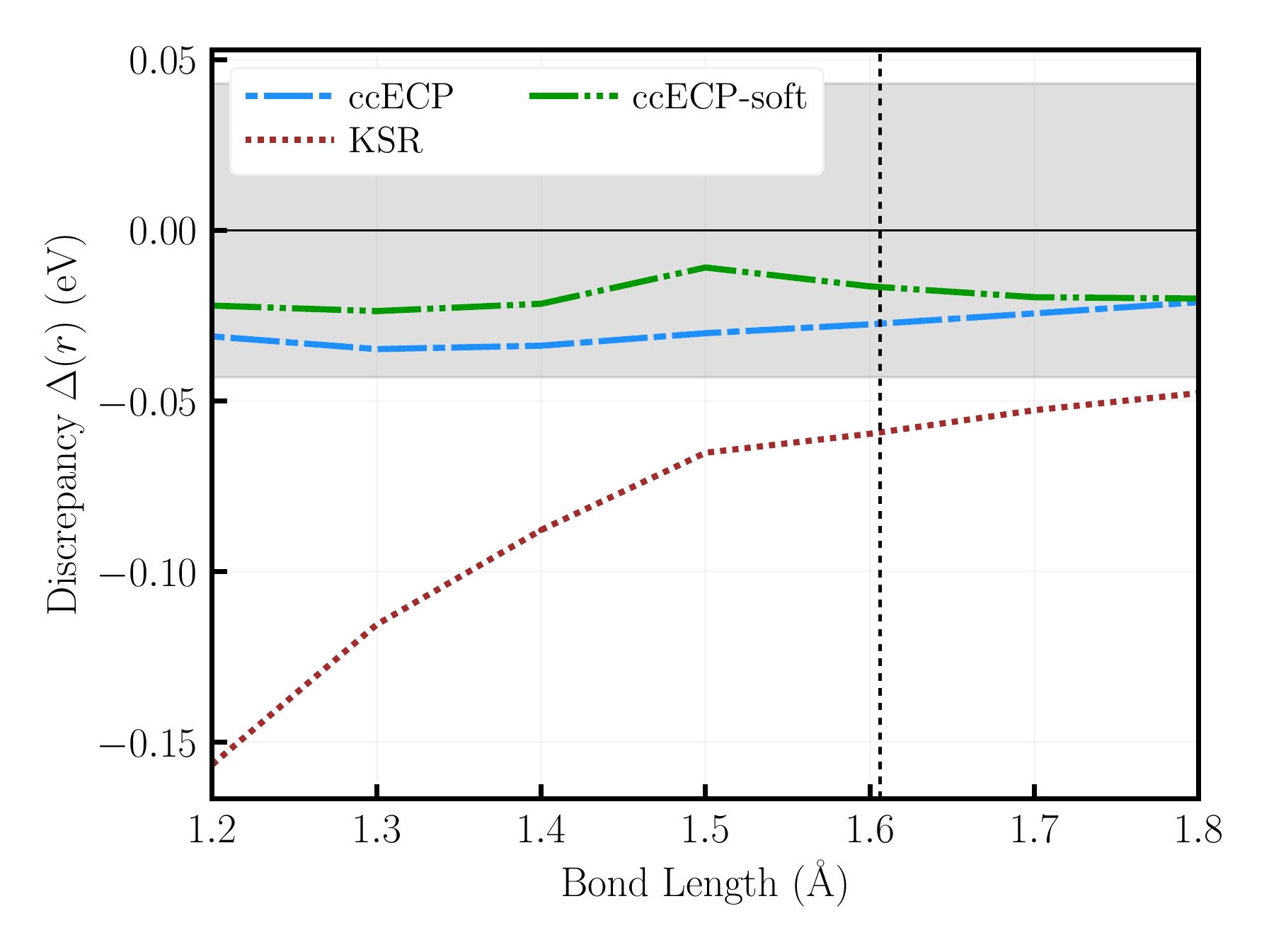}
\caption{FeO QZ binding curve discrepancies}
\label{fig:FeO}
\end{subfigure}
\caption{
Binding energy discrepancies for (a) FeH and (b) FeO molecules with the reference being the scalar relativistic all-electron CCSD(T) result. KSR denotes
the soft, DFT-derived ECP by Krogel, Santana and Reboredo
\cite{krogel_pseudopotentials_2016}.
The shaded region indicates the band of chemical accuracy. The dashed vertical line represents the equilibrium geometry.
}
\label{fig:Fe_mols}
\end{figure*}

\subsection{Co}
Similar results as for Fe have been obtained for Co. The atomic spectrum that included ionizations up to the $^5D(d^6)$ state came out marginally worse 
than for the original 
ccECP construction. However, both molecular binding curves show excellent agreement with the coupled cluster (CC) calculations within the chemical accuracy range. Reasonable plane wave energy cut-off  is approximately 360 Ry with the obvious caveat that this might differ somewhat depending on the system, accuracy thresholds and the used code. 

\begin{figure*}[!htbp]
\centering
\begin{subfigure}{0.5\textwidth}
\includegraphics[width=\textwidth]{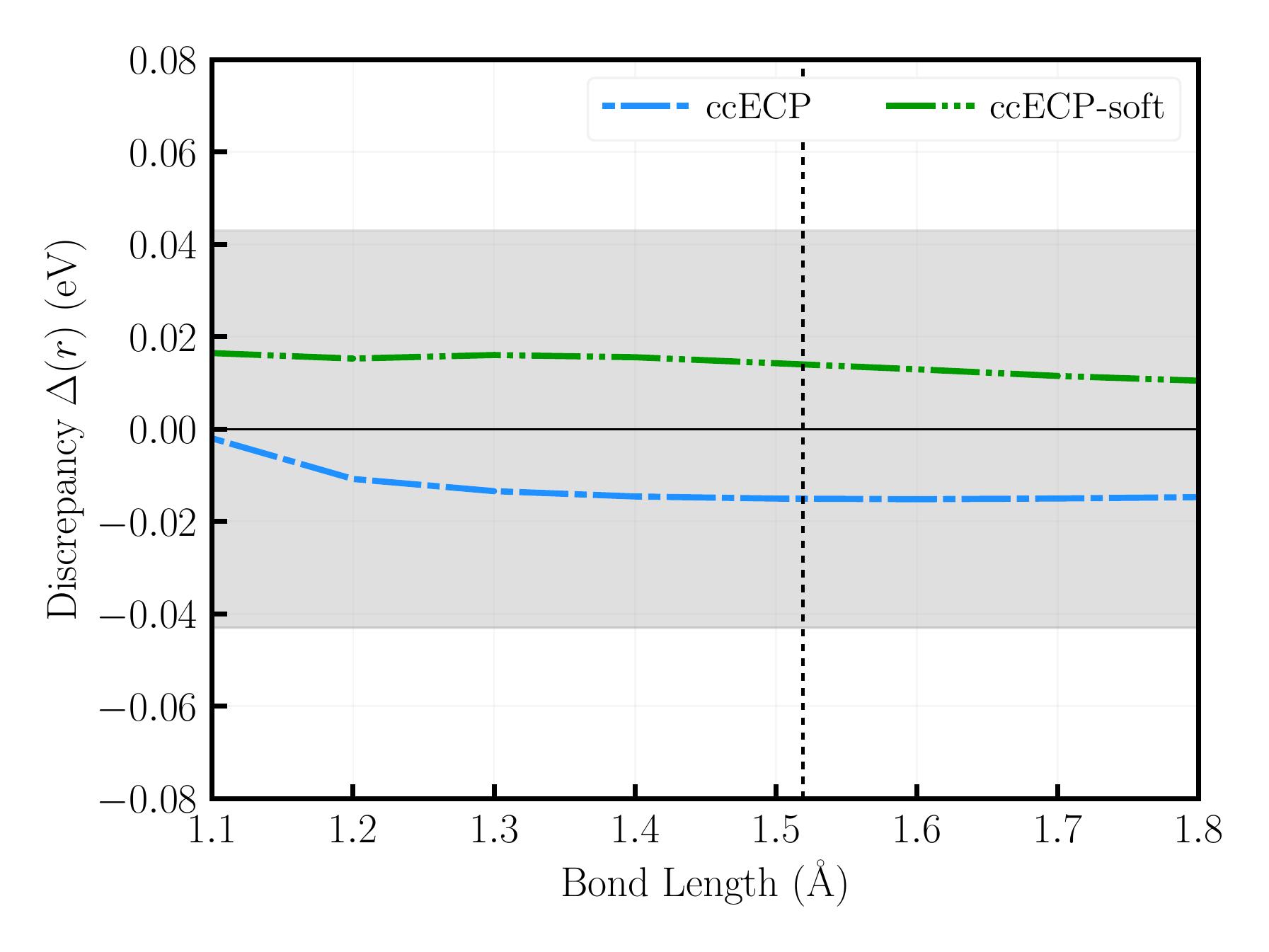}
\caption{CoH 5Z binding curve discrepancies}
\label{fig:CoH}
\end{subfigure}%
\begin{subfigure}{0.5\textwidth}
\includegraphics[width=\textwidth]{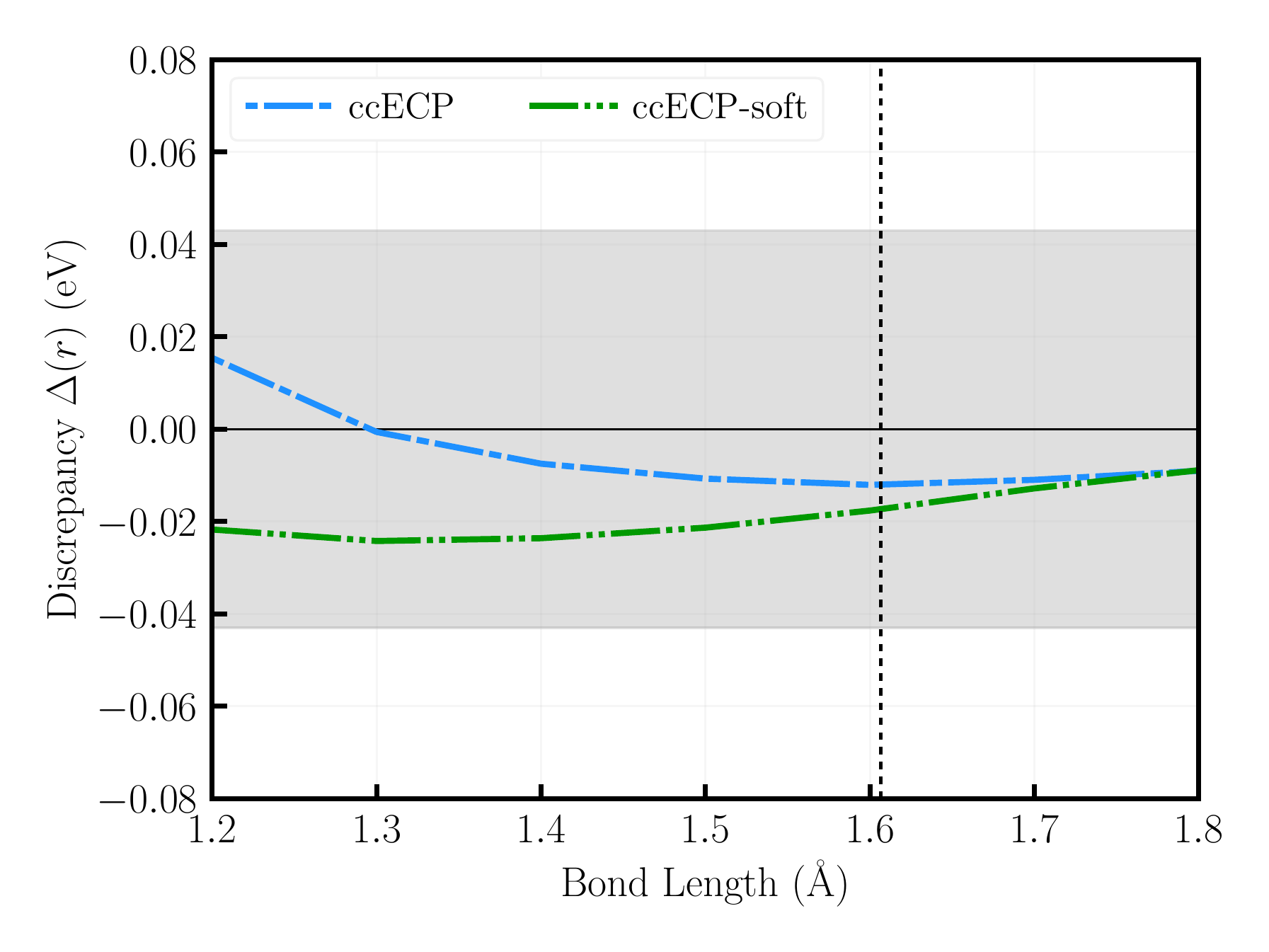}
\caption{CoO QZ binding curve discrepancies}
\label{fig:CoO}
\end{subfigure}
\caption{
Binding energy discrepancies for (a) CoH and (b) CoO molecules with the reference being the scalar relativistic all-electron CCSD(T) result.
The shaded region indicates the band of chemical accuracy. The dashed vertical line represents the equilibrium geometry.
}
\label{fig:Co_mols}
\end{figure*}

\subsection{Ni}
The results for the Ni atom show similar pattern apart from a minor accuracy compromise.
%Interestingly, the atomic spectrum that included ionizations up to the $^5D(d^4)$ state came out marginally  
%more accurate than for the %original ccECP construction. 
We observe that the hydride molecule bias is roughly constant along the whole binding curve. On the other hand, we clearly see small bias approaching the chemical accuracy at the very shortest bond lengths for the oxide dimer. We suspect that the  fidelity tuning is complicated by the well-known  $3d  \leftrightarrow 4s$ degeneracy/instability. Note, however, that around the bond equilibrium we see excellent agreement with all-electron reference and therefore we deem this minor accuracy compromise to be acceptable.
Probing such small bond lengths would correspond to extremely high pressures in solid systems such as NiO crystal. The plane wave cut-off is approximately 375 Ry based on atomic criteria in Opium, although  corresponding value in solid state calculations might differ somewhat depending on the system.

\begin{figure*}[!htbp]
\centering
\begin{subfigure}{0.5\textwidth}
\includegraphics[width=\textwidth]{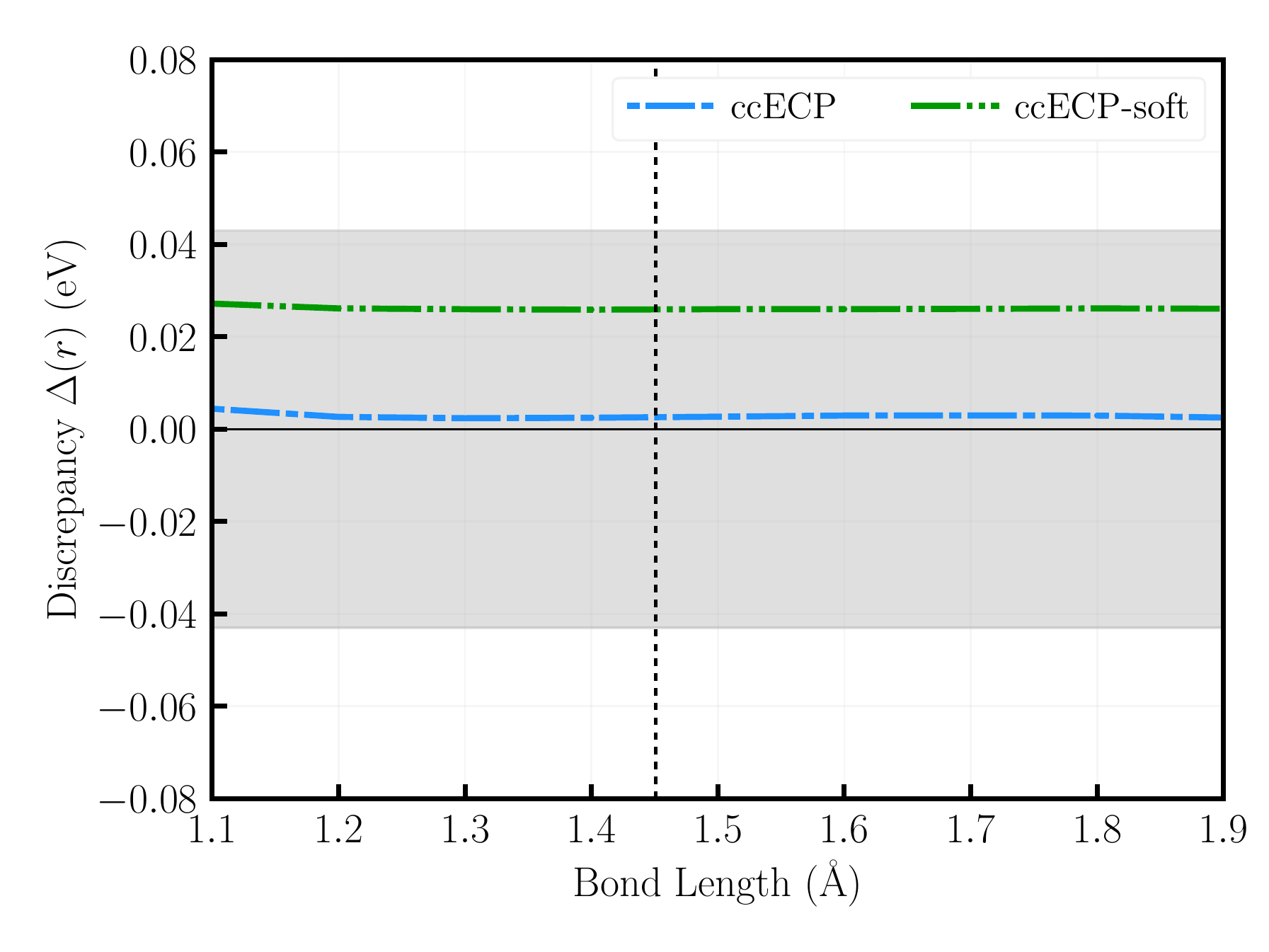}
\caption{NiH 5Z binding curve discrepancies}
\label{fig:NiH}
\end{subfigure}%
\begin{subfigure}{0.5\textwidth}
\includegraphics[width=\textwidth]{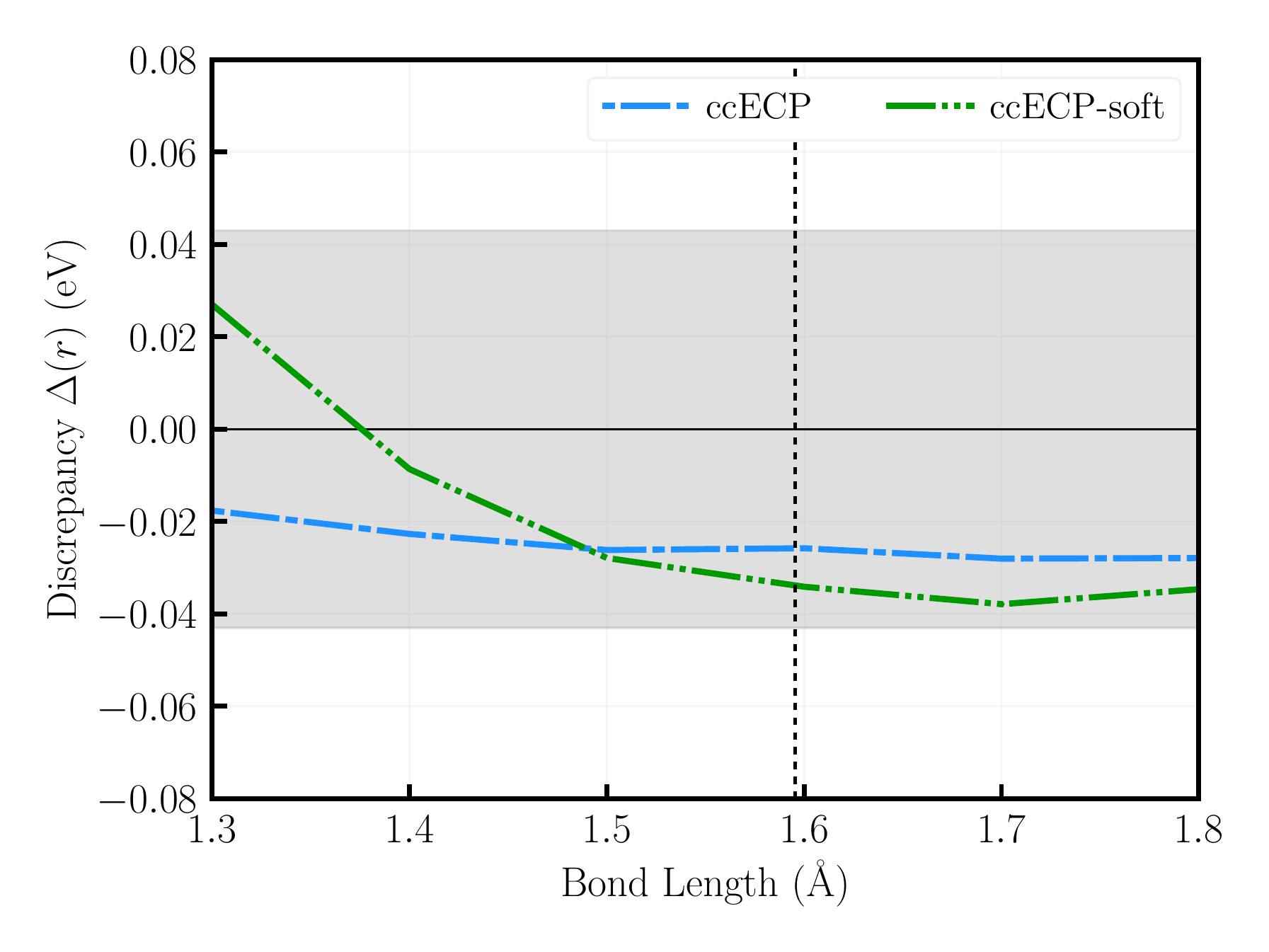}
\caption{NiO QZ binding curve discrepancies}
\label{fig:NiO}
\end{subfigure}
\caption{
Binding energy discrepancies for (a) NiH and (b) NiO molecules. The rest of notations is the same as in figures above. 
}
\label{fig:Ni_mols}
\end{figure*}

\subsection{Cu and Zn}

%\textit{ GM: The CuO graph at r=1.4 is not smooth due to difficult convergence issue. I am trying to rerun it.{\color{blue} LM: OK, good. That's minor.}}
Quality of ccECP-soft constructions for Cu and Zn atoms is comparable.
The requirement of smooth and as shallow as possible local potential is in contradiction with
the large number
of valence electrons and corresponding $Z_{\rm eff}$. 
As a consequence, 
the spectral errors are a bit larger but still acceptable. However, the molecular data shows essentially the same accuracy as the original ccECP construction. The approximate plane wave energy cut-off is about 400 Ry
which enables highly accurate calculations of solids and other systems in periodic setting. It is actually remarkable that even with rather deep semicore state  $3s$ one can achieve such a low cut-off within rather stringent accuracy requirements. 

\begin{figure*}[!htbp]
\centering
\begin{subfigure}{0.5\textwidth}
\includegraphics[width=\textwidth]{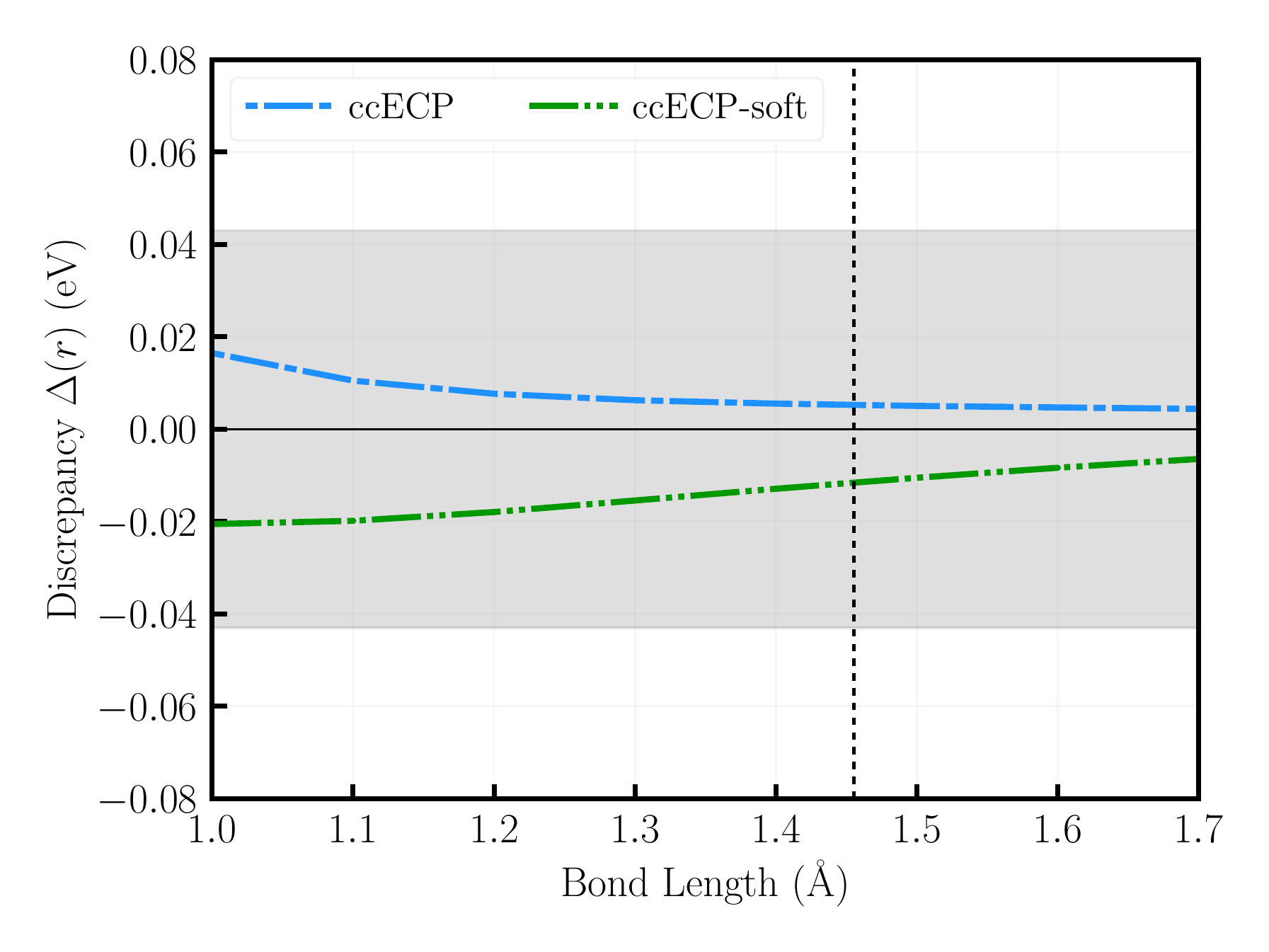}
\caption{CuH 5Z binding curve discrepancies}
\label{fig:CuH}
\end{subfigure}%
\begin{subfigure}{0.5\textwidth}
\includegraphics[width=\textwidth]{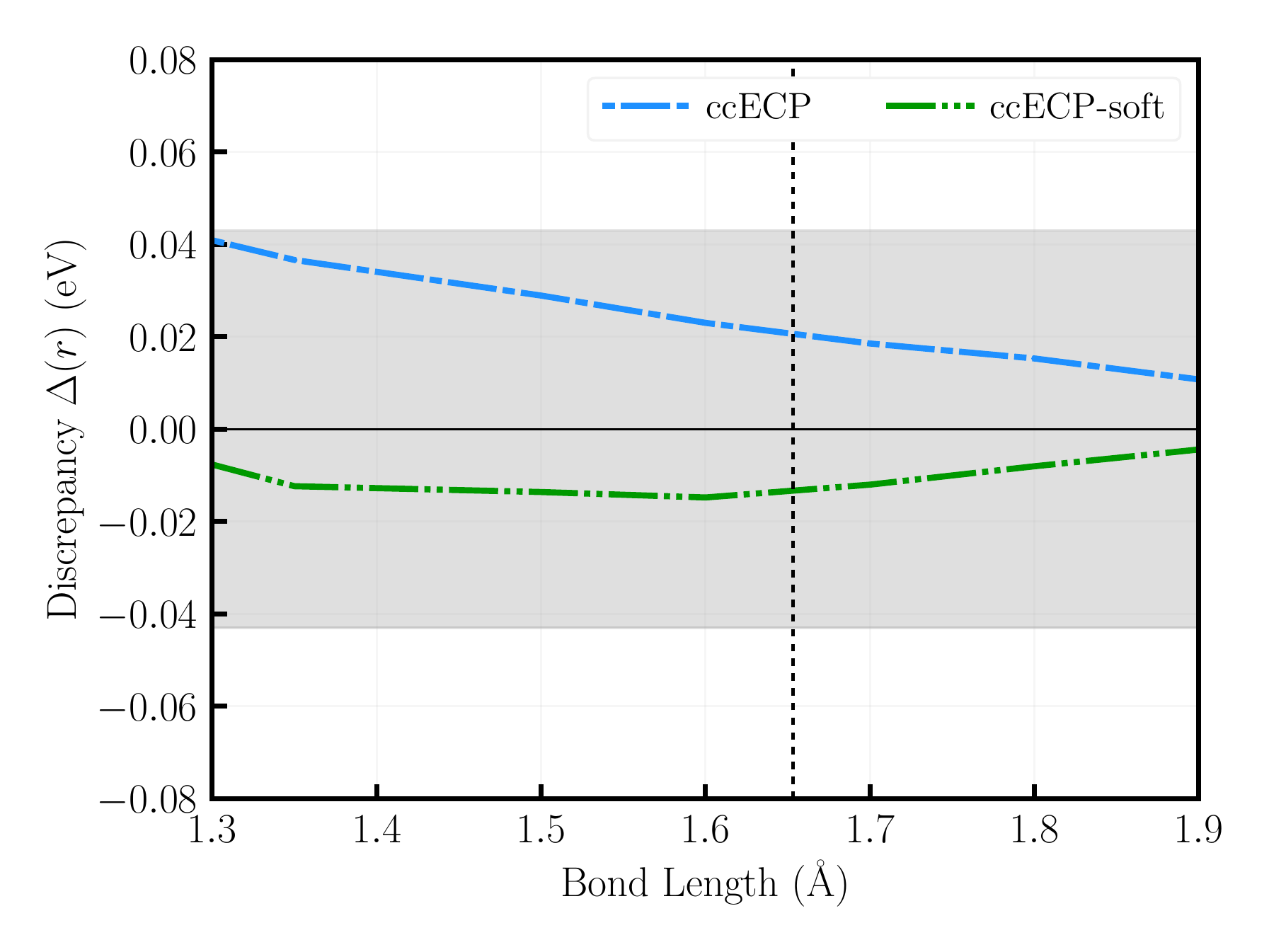}
\caption{CuO QZ binding curve discrepancies}
\label{fig:CuO}
\end{subfigure}
\caption{
Binding energy discrepancies for (a) CuH and (b) CuO molecules.
The rest of notations is the same as in figures above.
}
\label{fig:Cu_mols}
\end{figure*}

%\subsection{Zn}

\begin{figure*}[!htbp]
\centering
\begin{subfigure}{0.5\textwidth}
\includegraphics[width=\textwidth]{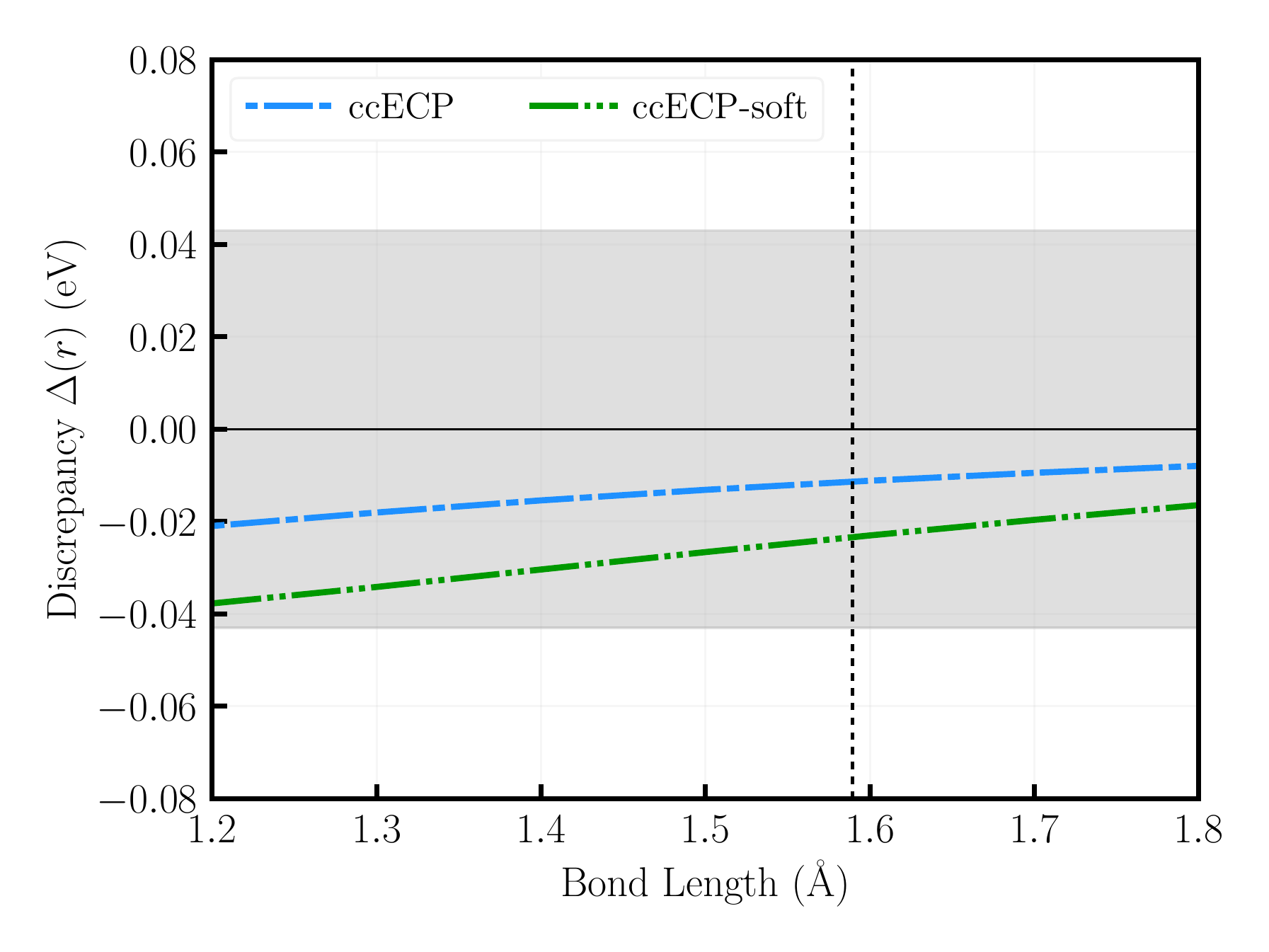}
\caption{ZnH 5Z binding curve discrepancies}
\label{fig:ZnH}
\end{subfigure}%
\begin{subfigure}{0.5\textwidth}
\includegraphics[width=\textwidth]{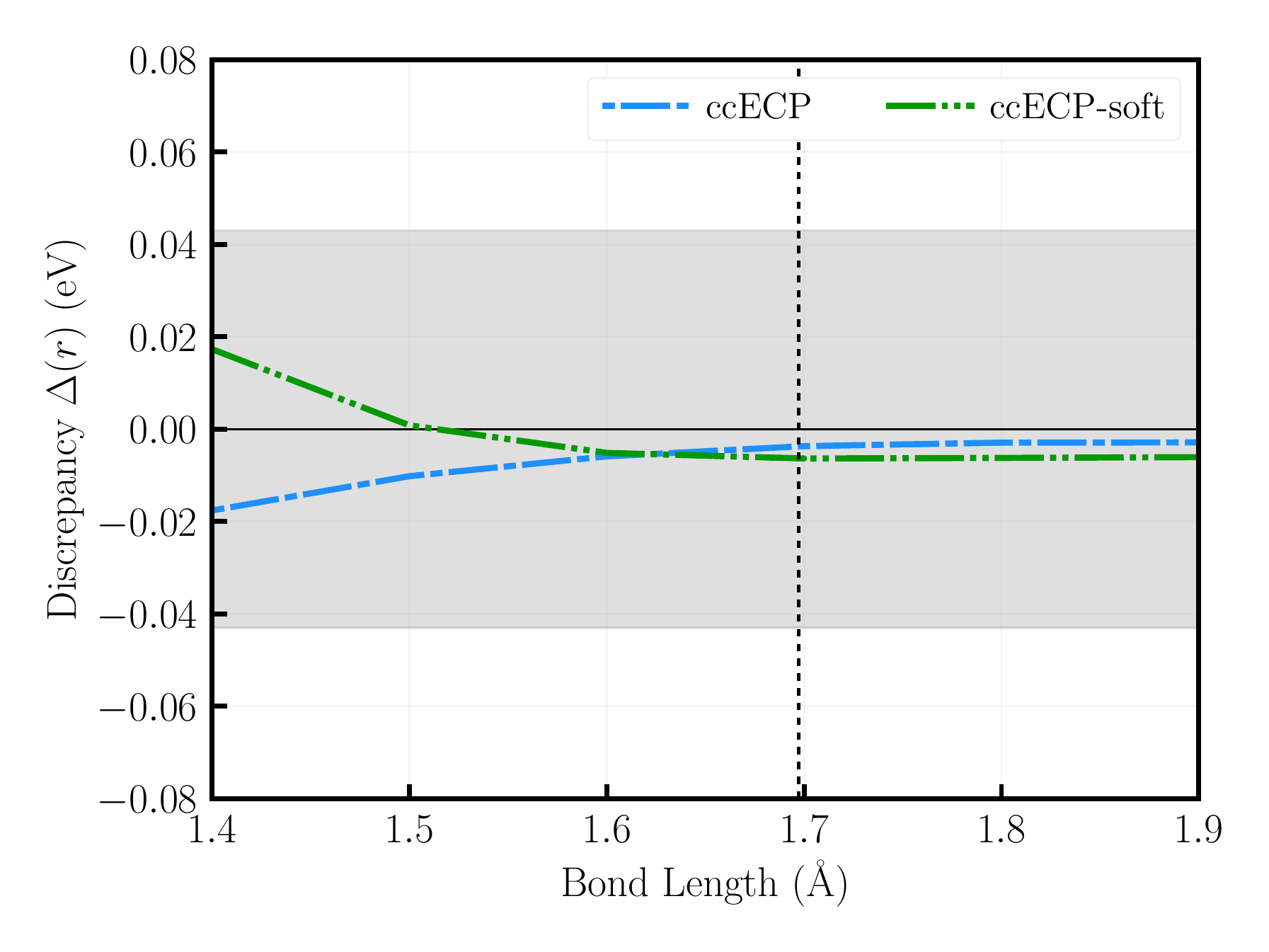}
\caption{ZnO QZ binding curve discrepancies}
\label{fig:ZnO}
\end{subfigure}
\caption{
Binding energy discrepancies for (a) ZnH and (b) ZnO molecules.
The rest of notations is the same as in figures above.
}
\label{fig:Zn_mols}
\end{figure*}

\subsection{Average molecular discrepancies}

In \tref{morse:all_ecps}, we present the summary of molecular binding property discrepancies for ccECP and ccECP-soft. Overall, we see comparable quality of both ECPs with all-electron binding parameters. This can be expected since the binding energy discrepancies figures for both ECPs are balanced within chemical accuracy for all the molecules with minor exception mentioned above for NiO, where ccECP-soft mildly underbinds at very compressed bond lengths.

\begin{table}[!htbp]
\centering
\caption{Mean absolute deviations of binding parameters for various core approximations with respect to AE correlated data for related hydride and oxide molecules. All parameters were obtained using Morse potential fit. The parameters shown are dissociation energy $D_e$, equilibrium bond length $r_e$, vibrational frequency $\omega_e$ and binding energy discrepancy at dissociation for compressed bond length $D_{diss}$.
}
\label{morse:all_ecps}
\begin{tabular}{l|rrrrrrrrrr}
\hline\hline
{} & $D_e$(eV) & $r_e$(\AA) & $\omega_e$(cm$^{-1}$) & $D_{diss}$(eV) \\
\hline
ccECP         &   0.0119(50) &  0.0008(17) &              2.8(5.5) &      0.015(47) \\
ccECP-soft    &   0.0174(50) &  0.0012(17) &              4.6(5.5) &      0.023(47) \\
\hline\hline
\end{tabular}
\end{table}

\subsection{Solid-state Performance and Cutoffs}
To verify the performance of the ccECP-soft series in planewave codes we ran several solid state calculations in QUANTUM ESPRESSO at various cutoffs to benchmark of the softened ccECPs.
We investigated FeO, CuO and ZnO 
in the LDA+U framework as a representative set for the series to determine the kinetic energy (KE) cutoffs for the ccECP-softs produced in this work.
Each system has a neutral unit cell, for FeO, the unit cell is rhombohedral with 2 Iron and 2 Oxygen atoms, and the solid is in an anti-ferromagnetic phase.
The unit cell of CuO studied here is in C2/c symmetry \cite{aasbrink1970refinement}, with 4 Copper and 4 Oxygen atoms; the solid is in the non-magnetic phase. 
Lastly, the unit cell of ZnO has a P63mc structure \cite{xu1993electronic}, with 2 Zinc and 2 Oxygen atoms. 
The solid is in the non-magnetic phase. 
%Each graph represents a number of runs constructed to determine the difference in total energy at various kinetic energy cutoffs. 
First a reference calculation is run at KE cutoff of $2000$ Ry, at which point the total energy should not meaningfully change if the cutoff is increased further, and beyond this value would become needlessly expensive to compute. 
Then a range of cutoffs are run from 300 Ry to 500 Ry in increments of 10 Ry to scan the discrepancy from the reference energy. 
The system is deemed "converged" with regard to cutoff, when the energy discrepancy per valence electron falls below 1 meV.%, indicated by the gray span near the x axis.
%The y-axis represents the total energy per chemical formula. 
Further discussion of atomic and molecular cutoff can be found in the supplementary materials. 
Lastly, we repeated the process described above for the original ccECP parameterizations from our previous work\cite{annaberdiyev_new_2018}, and found there cutoffs in a similar way for comparison purposes. 
The major difference was the amount of time each sub-calculation ran, and the cutoff at which the parameterizations converged. For the ccECP-softs the cutoffs for the TMOs we tested converged around 400 Ry, whereas for the ccECPs converged well after 1000 Ry. CuO even has a cutoff of 1500 Ry as seen in \tref{tab:tmo_cutoffs}. The graphs from which the table values were taken are included in the supplemental.
%\begin{figure*}[!htbp]
%\centering
%\begin{subfigure}{0.5\textwidth}
%\includegraphics[width=\textwidth]{figures/FeO_cutoff.pdf}
%\caption{FeO energy discrepancies at various cutoffs referenced to 2000 Ry result.}
%\label{fig:FeO}
%\end{subfigure}
%\\
%\begin{subfigure}{0.49\textwidth}
%\includegraphics[width=\textwidth]{figures/CuO_cutoff.pdf}
%\caption{CuO energy discrepancies at various cutoffs referenced to 2000 Ry result.}
%\label{fig:FeO}
%\end{subfigure}
%\begin{subfigure}{0.49\textwidth}
%\includegraphics[width=\textwidth]{figures/ZnO_cutoff.pdf}
%\caption{ZnO energy discrepancies at various cutoffs referenced to 2000 Ry result.}
%\label{fig:ZnO}
%\end{subfigure}
%\caption{}
%\label{fig:TMO_cutoffs}
%\end{figure*}

%\begin{figure*}[!htbp]
%\centering
%\begin{subfigure}{0.3\textwidth}
%\includegraphics[width=\textwidth]{figures/FeO_cutoff.pdf}
%\caption{FeO.}
%\label{fig:FeO}
%\end{subfigure}
%\begin{subfigure}{0.30\textwidth}
%\includegraphics[width=\textwidth]{figures/CuO_cutoff.pdf}
%\caption{CuO.}
%\label{fig:FeO}
%\end{subfigure}
%\begin{subfigure}{0.3\textwidth}
%\includegraphics[width=\textwidth]{figures/ZnO_cutoff.pdf}
%\caption{ZnO.}
%\label{fig:ZnO}
%\end{subfigure}
%\caption{FeO (a), CuO (b), and ZnO (c) energy discrepancies at various cutoffs referenced to their respective 2000Ry result.}
%\label{fig:TMO_cutoffs}
%\end{figure*}

\begin{table}[]
    \centering
    \begin{tabular}{l|cc}
    \hline\hline
        TMO & ccECP-soft(Ry) & ccECP(Ry) \\
        \hline
         FeO& 430 & 1140\\
         CuO& 390 & 1500\\
         ZnO& 390 & 1315\\
        \hline\hline
         
    \end{tabular}
    \caption{Comparison of cutoffs for several TMOs in the series using the newly created ccECP-soft and ccECP pseudopotentials. The old standard ccECPs generally triple the cutoff of the softened versions, leading to a much higher computational cost in this particular application.}
    \label{tab:tmo_cutoffs}
\end{table}

\newpage

\section{Conclusions}

We present a new modified set of
ccECP-soft for late $3d$ elements Mn-Zn with [Ne]-core
that can be used with plane wave codes. The two key goals we address are the accuracy through many-body construction and, at the same time, 
efficient application to plane wave codes with low energy cut-offs. For these purposes we use mildly less stringent criteria on spectral fidelity and on tests of molecular binding.  We do not include highly ionized states (beyond 6+) into the optimization and we allow smaller exponents in local channel that proved to be the key parameters for this purpose. This leads to smaller curvatures and more shallow local potentials in the core region. 

We obtained very encouraging results. In particular, for Cr-Ni
atoms the spectral errors are small and almost on par with our more stringent original construction.
Mildly less accurate spectra for Cu and Zn atoms are, however, still qualified for high accuracy calculations.  
With all the molecular binding energy discrepancy curves for the hydride and oxide molecules are strictly within the chemical accuracy showing excellent fidelity also in 
in bonding situations away from equilibrium. 
%It is very encouraging that the molecular curves are almost on par with the original constructions.
%We discvered only one minor deficiency
%for the very shortest bond of the NiO molecule.
Overall, the newly constructed 
and modified set of ccECP-soft psuedopotentials are basically in the same accuracy class as the original ccECPs.
We also expect essentially all the properties such as correlation energies calculated previously\cite{annaberdiyev_accurate_2020} to follow the same pattern with very minor differences, say, at few mHa level or so.

We believe that the constructed set will significantly expand the usefulness of ccECPs and provide tested and consistent
data for accurate valence-only electronic structure calculations in 
many-body approaches.

All ccECP-soft in various code formats 
and corresponding basis sets are available at \url{https://pseudopotentiallibrary.org}

%{\bf Add website, data }

\section{Acknowledgments}

We are  grateful to Paul R. C. Kent and Luke Shulenburger for reading the manuscript and helpful suggestions.

This work has been supported by the U.S. Department of Energy, Office of Science, Basic Energy Sciences, Materials Sciences and Engineering Division, as part of the Computational Materials Sciences Program and Center for Predictive Simulation of Functional Materials.

This research used resources of the National Energy Research Scientific Computing Center (NERSC), a U.S. Department of Energy Office of Science User Facility operated under Contract No. DE-AC02-05CH11231. 
This research used resources of the Argonne Leadership Computing Facility, which is a DOE Office of Science User Facility supported under contract DE-AC02-06CH11357. 
This research also used resources of the Oak Ridge Leadership Computing Facility, which is a DOE Office of Science User Facility supported under Contract DE-AC05-00OR22725.

%This paper describes objective technical results and analysis. Any subjective views or opinions that might be expressed in the paper do not necessarily represent the views of the U.S. Department of Energy or the United States Government.

\bibliography{main.bib}

\end{document}